\documentclass[prb,showpacs]{revtex4}
\usepackage{graphicx,psfrag,amssymb,amsmath,color}
\def\no{\noindent}
\def\bc{\begin{center}}
\def\ec{\end{center}}

\def\beq{\begin{equation}}
\def\eeq{\end{equation}}

\begin{document}
\title{Local density of states in disordered graphene}
\author{Klaus Ziegler}
\affiliation{Institut f\"ur Physik, Universit\"at Augsburg, D-86135 Augsburg, Germany}
\author{Bal\'azs D\'ora}
\email{dora@pks.mpg.de}
\affiliation{Max-Planck-Institut f\"ur Physik komplexer Systeme, N\"othnitzer Str. 38, 01187 
Dresden, Germany}
\author{Peter Thalmeier}
\affiliation{Max-Planck-Institut f\"ur Chemische Physik fester Stoffe, 01187 Dresden, Germany}
\date{\today}

\begin{abstract}

We study two lattice models, the honeycomb lattice (HCL) and a special square lattice (SQL), both reducing to the Dirac equation in 
the 
continuum 
limit. In the presence of disorder (gaussian potential disorder and random vector potential), we investigate the behaviour of the 
density of states (DOS) numerically and analytically.
While an upper bound can be derived for the DOS on the SQL at the Dirac point, which is also confirmed by numerical calculations,
no such upper limit exists on the HCL in the presence of random vector potential. A careful investigation of the lowest eigenvalues 
indeed indicate, that the DOS can possibly be divergent at the Dirac point on the HCL.
In spite of sharing a common continuum limit, these lattice models exhibit different behaviour.

\end{abstract}


\maketitle

\section{Introduction}

Graphene, a two-dimensional sheet of carbon atoms forming a honeycomb lattice,
has set the stage for studying Dirac-type quasiparticles in two dimensional materials
\cite{novoselov05,zhang05,geim07}. A substantial part of the investigation has been devoted to
the unusual transport properties of graphene. More recently, also local properties 
have been studied \cite{martin07,zhang08}.

Many physical properties depend directly or indirectly on the density of (quantum) states
at the Fermi energy. Therefore, the density of states (DOS), % $\rho(E)$ 
especially near the Fermi level, %(i.e. for $E\sim0$) 
is an interesting and important quantity to study. Local probing of graphene, 
such as in the recent STM experiments \cite{martin07,zhang08}, have
also raised interest in the local DOS. 
Moreover, the DOS at the Dirac point also plays an important role
as an indicator for spontaneous symmetry breaking, which causes
long-range correlations in graphene \cite{ziegler08}.

In pure graphene (or for pure Dirac fermions), in contrast to disordered graphene,
the DOS vanishes linearly like $\rho(E)\sim |E|$ at
the Dirac point $E=0$. Scattering by disorder may create new states at any energy, also at $E=0$.
As a consequence, the linear behavior of the DOS at low energies is affected by disorder.
On the other hand, the linear behavior of the DOS can be considered as a power law of a 
critical phenomenon with exponent 1. In fact, the
phase transition in the 2D Ising model is directly linked to this linear behavior of the
DOS of 2D Dirac fermions \cite{schultz64}. 
A common belief is that disorder or additional interaction effects do not
destroy the critical phenomenon but only modify the exponent of the corresponding power law. 
This possibility has also been discussed for the Dirac fermions, for instance,
in the case of a random vector potential \cite{ludwig94,nersesyan94,dora08}. 
Another possibility is that disorder creates a new intermediate phase between the two
phases of the pure system \cite{griffiths69}.

For weak disorder we can apply a perturbation theory with respect to a random vector potential. 
This approach gives a power law   
\beq
\langle\rho(E)\rangle\sim |E|^\alpha \ \ \ (\alpha\le 1) \ ,
\label{dos01}
\eeq
where the exponent decreases with increasing variance of the disorder distribution $g$ as 
\beq
\alpha \sim 1-g/\pi \ .
\label{exponent}
\eeq
On the other hand,
there has been a long debate in the literature whether or not the exponent can have
negative values for strong disorder (i.e., whether or not there is a divergent average DOS in the case of
strong disorder) for the model with a single
Dirac cone \cite{ludwig94,nersesyan94,altland02,guinea08}.

The case of two Dirac cones with intervalley scattering has also been discussed intensively in the 
literature \cite{nersesyan94,caux98,bhaseen01,foster08}.
Intervalley scattering may affect the density of states strongly, leading to a power law
with a universal exponent $\alpha=1/7$ for any strength of disorder \cite{nersesyan94}.

The power law of the density of states has direct implications for the
transport properties. The Einstein relation states that the conductivity $\sigma$ and the DOS are 
proportional to each other:
\[
\sigma\propto \rho (E) D(E) \ ,
\]
where $D(E)$ is the diffusion coefficient. If $\rho(E)$ vanishes at the Dirac point $E=0$ for $\alpha>0$, 
the conductivity also vanishes, as long as $D(E=0)$ is finite. The latter should be the case in the presence
of disorder because $D(E)$ measures the amount of scattering, since $D$ is proportional to the scattering
time $\tau$.
An exceptional case is a pure system, where transport is ballistic ($D(E\rightarrow 0)\rightarrow\infty$). 
On the other hand, if $\rho(E)$ diverges at the Dirac point for $\alpha<0$, the conductivity also diverges, 
unless the diffusion coefficient vanishes. 

An alternative approach for the density of states is the self-consistent non-crossing 
(or Born) approximation \cite{suzuura02,peres06,koshino06}.
The perturbative result of the DOS in Eqs. (\ref{dos01}), (\ref{exponent}) was confirmed for the 
tight-binding model on the honeycomb lattice within the self-consistent calculation \cite{dora08}. 
However, very close to $E=0$ an interception of the power law was found, indicating a non-zero DOS 
at $E=0$. Moreover, the calculation gave only positive exponents $\alpha$, even for strong disorder, 
in contrast to the exponent suggested in Ref. \cite{ludwig94,nersesyan94}
\[
\alpha =\frac{1-g/\pi}{1+g/\pi} \ .
\]

In order to shed some light on the behavior of the average DOS near the Dirac point,
we shall focus in this paper on two cases: (i) a single Dirac cone with random vector potential
and (ii) the honeycomb lattice with unidirectional random bonds. By comparing these two cases
we estimate the effect of intervalley scattering on the DOS.
%The single Dirac cone by projecting the tight-binding model of the honeycomb lattice
%onto one of its Dirac cones.

The paper is organized as follows: After a brief introduction of the tight-binding model
for graphene and the projection to a single Dirac cone we discuss the underlying symmetries
of the models in Sect. 2. Based on these considerations we derive a simple expression for the local DOS
in the case of the single Dirac cone with random vector potential in Sect. 2.2. 
This allows us in Sect. 2.3 to give an upper bound for the average local DOS.
In the second part of the paper (Sect. 3) we apply exact diagonalization to the single Dirac cone 
with random vector potential and to the tight-binding model on the honeycomb lattice with 
unidirectional bond  disorder to study the energy levels near the Dirac point for finite systems.

\section{Models and symmetries}

Starting point is a tight-binding model for quasiparticles on the honeycomb lattice.
The honeycomb lattice is a bipartite lattice. After dividing it into sublattice A and B, 
the quasiparticles are pseudospin-1/2 particles with respect to the two sublattices,
and the corresponding Hamiltonian has a chiral symmetry. This allows us to write
\[
{\bf H}=\sum_{r,r'}\sum_{j,j'=1,2}H^{\rm HCL}_{r,j;r',j'}c^\dagger_{r,j}c_{r',j'} \ ,
\]
where $r$ runs over sublattice A and $j$ refers to sublattice A ($j=1$)
and sublattice B ($j=2$). The only energy scale of this Hamiltonian is the hopping energy $t$.
Then the Hamiltonian matrix can be expressed with Pauli matrices as  \cite{ziegler06}
% H=H_0+V\sigma_1 +V'\sigma_2\ ,
% \hskip0.5cm
\beq
H^{\rm HCL}=h_1\sigma_1+h_2\sigma_2 \ .
\label{ham00}
\eeq
$h_1$ and $h_2$, defined on sublattice A, are symmetric and antisymmetric matrices
($h_1^T=h_1$, $h_2^T=-h_2$), respectively. The off-diagonal element of the Pauli matrices
connect the two sublattices. These properties imply a real symmetric Hamiltonian. %: $H^T=H$. 
The corresponding quasiparticle dispersion has
two Dirac cones (two ``valleys'') at low energies.

\subsection{Dirac Hamiltonian}

\begin{figure}[h!]
\centering
\psfrag{t1}[][][1][0]{$it$}
\psfrag{t2}[][][1][0]{$-t$}
\psfrag{t3}[][][1][0]{$t$}
{\includegraphics[width=4cm,height=4cm]{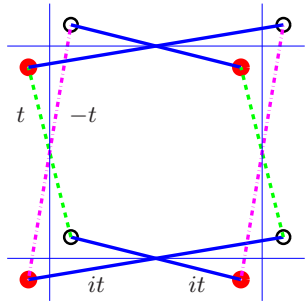}}

\caption{(Color online) The square lattice, whose continuum limit is the Dirac Hamiltonian is visualized. Filled red and empty black 
circles denote up and down spins at a given lattice point, thick/thin lines denote the hopping/lattice. The hopping matrix elements 
are indicated. Note the spin dependent hopping amplitudes!}
\label{sqham}
\end{figure}

Considering quasiparticles at
low energies only, we can expand the Hamiltonian around both Dirac points. Then we get a model 
that describes two separate spin-1/2 Dirac spinors. Scattering by disorder can, in principle, 
connect these two Dirac cones (valleys). It has been discussed that this leads to the $SU(2)$ 
Wess-Zumino-Witten model \cite{bhaseen01} (but see also Ref. \cite{altland02}). 
%However, in the absence of parity breaking there should be no Wess-Zumino-Witten term \cite{altland02}.
On the other hand, if inter-cone scattering 
is ignored (for instance, by assuming a smooth scattering potential that is constant
on the scale of the lattice spacing), the two valleys of the model are completely isolated from each
other and each valley can be studied separately. Then disorder can  
appear as a random scalar potential, a random mass or a random vector potential 
\cite{ludwig94}. Only the latter preserves the continuous chiral symmetry. It is 
believed that this type of disorder is related to ripples in the graphene sheet 
\cite{morozov06,castroneto07b}. The corresponding Hamiltonian $H_D$ is again a chiral 
spinor-1/2 Hamiltonian but, in contrast to the real symmetric tight-binding Hamiltonian on the 
honeycomb lattice $H_{\rm HCL}$, it breaks the time-reversal invariance
\beq
% H=H_0+V\sigma_1 +V'\sigma_2\ ,
% \hskip0.5cm
H_D=h_1\sigma_1+h_2\sigma_2 \ .
\label{ham0}
\eeq
$h_1$ and $h_2$ are now antisymmetric spatial matrices ($h_j^T=-h_j$ ($j=1,2)$) with imaginary
matrix elements, and $\sigma$ here denotes the physical spin, this is why this Hamiltonian breaks the time 
reversal invariance.  This gives $H_D^*=\sigma_2 H_D\sigma_2$.
Moreover, we assume that $h_j$ are lattice hopping matrix elements with nearest-neighbor
elements on a square lattice whose continuum limit is the $j$ component of the 2D gradient $\nabla_j$. 
This fictitious square lattice is sketched in Fig. \ref{sqham} with  spin dependent hopping amplitudes.
Thus, the Hamiltonian $H_D$ describes lattice Dirac fermions. The lattice constant is not that of
the original honeycomb lattice of the graphene sheet but larger, and related to 
the projection onto a single Dirac cone. In this respect the lattice structure of $H_D$ 
corresponds with the network approximation of the honeycomb lattice \cite{snyman08}.

In the following, disorder due to ripples will be considered. This can be represented by a random vector 
potential $(V_{1,r},V_{2,r})$ as
\beq
H=(h_1+V_1)\sigma_1+(h_2+V_2)\sigma_2 \ .
\label{ham5}
\eeq
% (or for Dirac fermions: $h_j^T=-h_j$) %, commuting matrices $[h_1,h_2]=0$,
% and random vector potentials $V$, $V'$ 
This Hamiltonian has three essential symmetry properties: It is Hermitian 
(i.e. $H^\dagger=H$), and it satisfies the following relations:
\beq
\sigma_3 H\sigma_3 = -H \ ,
\label{symm1}
\eeq
and with the staggered diagonal matrix $D$
\[
D_{rj,r'j'}=(-1)^{r_1+r_2}\delta_{r,r'}\delta_{j,j'}
\]
we get (cf. Appendix A)
\beq
\sigma_1 D H^T D\sigma_1 = H \ .
\label{symm2}
\eeq
The fact that $H$ is Hermitian implies for the Green's function 
$
G(i\epsilon)=(i\epsilon+H)^{-1}
$
the relation
\beq
G^\dagger(i\epsilon)=G(-i\epsilon) \ .
\label{symm1b}
\eeq
Moreover, Eq. (\ref{symm1}) implies
\beq
\sigma_3G(i\epsilon)\sigma_3=-G(-i\epsilon) \ ,
\label{symm2b}
\eeq
and Eq. (\ref{symm2}) implies
\beq
\sigma_1DG(i\epsilon)^TD\sigma_1=G(i\epsilon) \ .
\label{symm3b}
\eeq
The spatial diagonal elements of the Green's function $G_{rr}(i\epsilon)$ can be 
expressed in terms of Pauli matrices as
\beq
G_{rr}(i\epsilon)=g_0(i\epsilon)\sigma_0 %+g_1(i\epsilon)\sigma_1
+g_1(i\epsilon)\sigma_1+g_2(i\epsilon)\sigma_2 +g_3(i\epsilon)\sigma_3\ .
\label{expans}
\eeq
The three relations in Eqs. (\ref{symm1b}) -- (\ref{symm3b}) provide the following 
relations between the coefficients of the Pauli matrices:
\[
g_0^*(i\epsilon)=g_0(-i\epsilon)=-g_0(i\epsilon) \ ,
\hskip0.5cm
g_1^*(i\epsilon)=g_1(-i\epsilon)=g_1(i\epsilon) \ ,
\hskip0.5cm
g_2^*(i\epsilon)=g_2(-i\epsilon)=g_2(i\epsilon) \ , % \ , \hskip0.5cm
\]
\[
g_3(i\epsilon)=0 \ .% g_3(-i\epsilon)=-g_3(i\epsilon) \ .
\]
Note, that this is a clear consequence of Eq. \eqref{symm3b}, which holds true only on the square lattice.
Thus, $g_0$ is purely imaginary, whereas $g_1$ and $g_2$ are real and $g_3$ vanishes:
\beq
G_{rr}(i\epsilon)=g_0(i\epsilon)\sigma_0 %+g_1(i\epsilon)\sigma_1
+g_1(i\epsilon)\sigma_1+g_2(i\epsilon)\sigma_2\ .
\label{expansa}
\eeq

\subsection{Local density of states of Dirac fermions}

The Green's function $G=(i\epsilon+H_D)^{-1}$ allows us to write for the local DOS for a fixed random disorder configuration
\beq
\rho_r=-\frac{1}{2\pi} \textmd{ImTr}_2(G_{rr}) \ ,
\label{dos0}
\eeq
where $\epsilon>0$ is implicitly sent to zero, and the Tr is taken over the Pauli matrices. 
As a function of the random vector potential at site $r$ $(V_{1,r}, V_{2,r})$, the local DOS
$\rho_r$ of the Green's function in Eq. (\ref{expansa}) has a Lorentzian form 
(cf. Eq. (\ref{gf1}) in Appendix B):
\beq
\rho_r % =\frac{1}{2\pi} \textmd{Im Tr}_2(G_{rr}) 
= \frac{1}{\pi}\frac{(X_0+\epsilon)}{(X_0+\epsilon)^2+(X_1+V_{1,r})^2+(X_2+V_{2,r})^2}
\label{dos1}
\eeq
with some real variables $X_1,X_2$ and a positive real variable $\epsilon+X_0$, where the latter
is proportional to $\epsilon$. They depend
on $V_{1,r'}, V_{2,r'}$ for $r'\ne r$ but not on $V_{1,r}, V_{2,r}$. This expression can also be used to
determine the DOS away from the Dirac point at energy $E\ne0$ by replacing $\epsilon\to\epsilon-iE$:
\beq
\rho_r(E) = 
\frac{1}{\pi}\textmd{Re}\left[\frac{(X_0+\epsilon-iE)}{(X_0+\epsilon-iE)^2+(X_1+V_{1,r})^2+(X_2+V_{2,r})^2}
\right] \ .
\label{dos2}
\eeq
It should be noticed that this form of the local DOS is very special for the Green's function in 
Eq. (\ref{expansa}).
For instance, we would not get a Lorentzian in the case of a random scalar potential.

\begin{figure}[h!]
\centering
\psfrag{x}[t][b][1][0]{$\omega/D$}
\psfrag{y}[b][t][1][0]{$\langle\rho_r(E)\rangle$}
{\includegraphics[width=6cm,height=5cm]{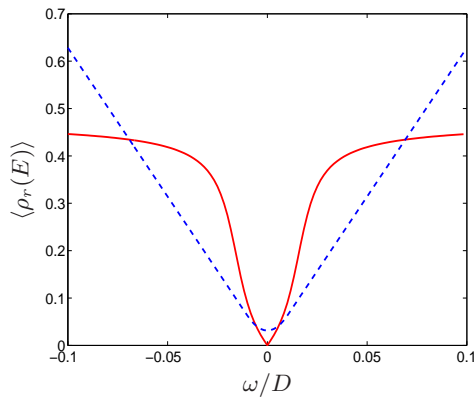}}

\caption{(Color online) Average local DOS $\langle\rho_r(E)\rangle$ of the Dirac Hamiltonian (Eq. (\ref{ham0})) 
for a local random vector potential $V_{1,r}$ (red curve) and a local random scalar potential $E_r$ (blue dashed curve). 
The potentials are box distributed with $-1\le V_{1,r}\le 1$ and $-0.1\le E_r\le 0.1$, $D$ is the cutoff in the continuum theory.
}
\label{dosfig0}
\end{figure}

Expression (\ref{dos2}) enables us to evaluate the local DOS $\rho_r(E)$
for an impurity at site $r$.  %$\langle\rho_r(E)\rangle$
According to Eq. (\ref{xparameters}) the parameters $X_j$ ($j=0,1,2,3$) of the system
without disorder are
\[
X_0=-\epsilon+iE+i\frac{g_0}{g_1^2-g_0^2},\ \ \ X_1= %\frac{g_1}{g_1^2-g_0^2},\ \ \ 
X_2=X_3=0 \ ,
\] 
where 
\beq
g_0=-\int\frac{i\epsilon+E}{(\epsilon-iE)^2+k^2}\frac{d^2k}{(2\pi)^2}.
% \ \ g_1=\int\frac{k_1+\kappa}{(\epsilon-iE)^2+(k_1+\kappa)^2+k_2^2}\frac{d^2k}{(2\pi)^2} \ .
\label{gnot}
\eeq
The local DOS of Eq. (\ref{dos2}) then reads
\[
\rho_r(E) = 
\frac{1}{\pi}\textmd{Re}\left[\frac{ig_0}{(1+g_1 V_{1,r})^2-g_0^2V_{1,r}^2}
\right] \ .
\]
We can also study a local scalar potential $E_r$ by adding the latter to the energy $E$
in $g_0$ of Eq. (\ref{gnot}). The contribution of the local potentials $E_r$ and $V_{1,r}$
to $\langle\rho_r(E)\rangle$ is quite different, as shown in Fig. \ref{dosfig0}. While the
scalar potential creates mostly states at and very close to the Dirac point, the vector potential
creates states in some distance from the Dirac point.

A direct evaluation of the variables $X_j$ ($j=0,1,2,3$) is difficult in the general case, where
we have a random vector potential at all sites. However, for finite and sufficiently small systems
an exact diagonalization is possible. Moreover, we can derive an upper bound for the average local DOS.
This will be discussed in the next section.

\subsection{Upper bound for the DOS of Dirac fermions}

Now we perform the integration with respect to $(V_{1},V_{2})$ for all sites to evaluate
the average local DOS.  For simplicity, we consider only the Dirac point $E=0$ here:
\beq
\langle \rho_{r}\rangle 
=\int\rho_{r}\prod_{r'}P(V_{1,r'})dV_{1,r'}P(V_{2,r'})dV_{2,r'} \ .
\label{aldos}
\eeq
First, we perform the integration with respect to $V_{1,r}$, using the expression
of $\rho_r$ in Eq. (\ref{dos1})
\[
\int\rho_r P(V_{1,r})dV_{1,r}=
\frac{1}{\pi}\int
\frac{(X_0+\epsilon)}{(X_0+\epsilon)^2+(X_1+V_{1,r})^2+(X_2+V_{2,r})^2}P(V_{1,r})dV_{1,r} \ .
\]
An upper bound for this integral is obtained from pulling out the maximum of the
distribution density $P(V_{1,r})$ which we call $P_m$: $P(V_{1,r})\le P_m$.
This gives 
\[
\int\rho_r P(V_{1,r})dV_{1,r}\le  \frac{P_m}{\pi}\int
\frac{(X_0+\epsilon)}{(X_0+\epsilon)^2+(X_1+V_{1,r})^2+(X_2+V_{2,r})^2}dV_{1,r} \ ,
\]
and after integrating over the Lorentzian function, which gives $\pi$, the right-hand side
becomes $P_m$:
\[
\int\rho_r P(V_{1,r})dV_{1,r}\le P_m \ .
\]
Going back to the expression in Eq. (\ref{aldos}), we obtain
\[
\langle \rho_{r}\rangle \le
P_m\int P(V_{2,r})dV_{2,r} \int \prod_{r'\ne r}P(V_{1,r'})dV_{1,r'}P(V_{2,r'})dV_{2,r'} =P_m \ .
\]
In other words, the averaged local DOS at the Dirac point $E=0$ has an upper bound:
\beq
\langle \rho_{r}\rangle =\frac{1}{2\pi} Tr_2(\langle \textmd{Im} G_{rr}\rangle ) 
%\le\int P_m\max_{V_{r'1},V_{r'2},r'\ne r}\int_{-\infty}^\infty\rho_{r}dV_r\prod_{r'}P(V_r{r'}')dV_{r'}'
\le \max_{-\infty<V<\infty}P(V) \ .
\label{upperbound}
\eeq
This means that for any smooth bounded distribution of $V_{1,r}$ (e.g. for a Gaussian) 
the corresponding average local DOS $\rho_r$ is finite. For discrete distributions, such as a
binary alloy, the upper bound is infinite though.

\section{Exact diagonalization}

For a better understanding of the details of the DOS,
we employ an exact diagonalization study on small clusters for both models, the Hamiltonian of Eq. (\ref{ham00}) 
on the original honeycomb lattice (HCL) and the Hamiltonian of Eq. (\ref{ham5}) on the effective square lattice (SQL).
Although both models reduce to the same continuum limit of Dirac fermions with random vector potential, 
they possess distinct structures in the DOS, as we will discuss below. We use Gaussian disorder with standard 
deviation $V$ (i.e., $V^2$ is the variance).

\subsubsection{Density of states by ED}

\begin{figure}[h!]
\centering
\psfrag{x}[t][b][1][0]{$\omega/t$}
\psfrag{y}[b][t][1][0]{$t\rho(\omega)$}
\psfrag{xx}[b][][1][0]{$\omega/t$}
\psfrag{yy}[][][1][0]{$t\rho(\omega)$}
\psfrag{uni}[l][][1][0]{ED, unidirectional}
\psfrag{pot}[l][][1][0]{ED, potential}
\psfrag{nca}[l][][1][0]{SCNCA}
{\includegraphics[width=14cm,height=7cm]{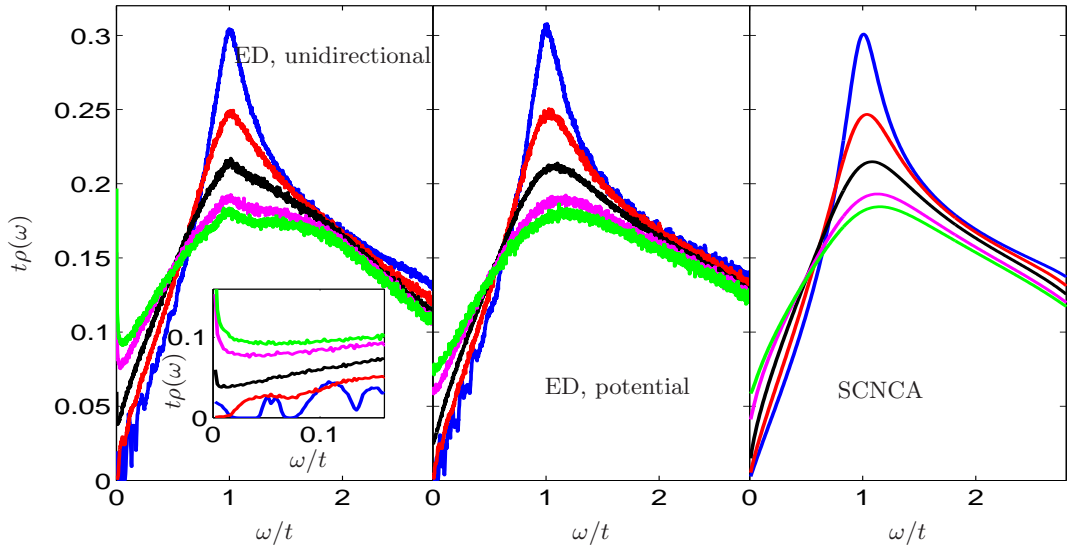}}

\caption{(Color online) The DOS is shown as obtained by exact diagonalization on $100\times100$ honeycomb clusters with Gaussian
unidirectional bond disorder (left panel), potential disorder (middle panel) after 1000 averages for $V_1/t=0.3$ (blue), 0.5 (red),
0.7 (black), 0.9 (magenta) and
1 (green). The right panel shows the corresponding self-consistent non-crossing approximation for the same parameters for the HCL. 
The inset shows the narrow peak at the DP for the unidirectional case. The SCNCA leads to the same result for pure unidirectional 
bond or potential disorder. Note the nice agreement between the numerical and analytical results for weak disorder! }
\label{dosedvsnca}
\end{figure}

Determining the DOS of the infinite system by studying a finite system is a difficult task, since any finite 
system possesses distinct energy levels, resulting in separate Dirac delta peaks in the DOS at the quasiparticle energies.
The DOS becomes continuous only in the thermodynamic limit.
In order to avoid this problem, we choose an indirect approach to evaluate the DOS by counting the number of eigenvalues 
in a narrow frequency range around a given energy $E$.
Strictly speaking, this leads to the number of states around $E$, but if the DOS is a smooth function,
this provides us with a sensible definition.
We obtain the DOS shown in Fig. \ref{dosedvsnca} on a $100\times 100$ HCL cluster with periodic boundary conditions for 
unidirectional bond and potential disorder, using a $t/500$ wide 
energy windows, where $t$ is the uniform hopping amplitude. For comparison, we also show the result of the 
self-consistent non-crossing approximation (SCNCA) on the HCL \cite{dora08}. As is seen, the agreement is surprisingly good 
for weak disorder, except for the case of bond disorder in a very close vicinity of the Dirac point. 
There, for $V_1 \lesssim 0.6t$, the residual DOS remains zero, which is in contrast
to the finite, although exponentially small, residual value for the case of potential disorder, described correctly by the SCNCA.
A narrow peak appears at the Dirac point (DP) for bond disorder if $V_1 \gtrsim 0.6t$. 
Whether this peak remains finite or diverges cannot be decided within this calculation of the DOS.
It should be mentioned that the DOS on a SQL is qualitatively similar to the potential disorder case on a HCL for strong disorder.
In particular, it never diverges at the DP.  
The anomalous behavior close to the DP is obvious in perturbation theory as well\cite{dora08}, where a dynamically 
generated low energy scale, similar to the Kondo scale, separates the high and low energy regions in the DOS. 

\subsubsection{Eigenvalues}

\begin{figure}[h!]
\centering
\psfrag{x}[t][b][1][0]{$V_1/t$}
\psfrag{y}[b][t][1][0]{100 smallest eigenvalues}
\psfrag{xx}[t][b][0.8][0]{$V_1/t$}
\psfrag{yy}[][][0.8][0]{smallest eigenvalues}

\psfrag{1000x1000}[][][1][0]{1000x1000 HCL}
{\includegraphics[width=10cm,height=7cm]{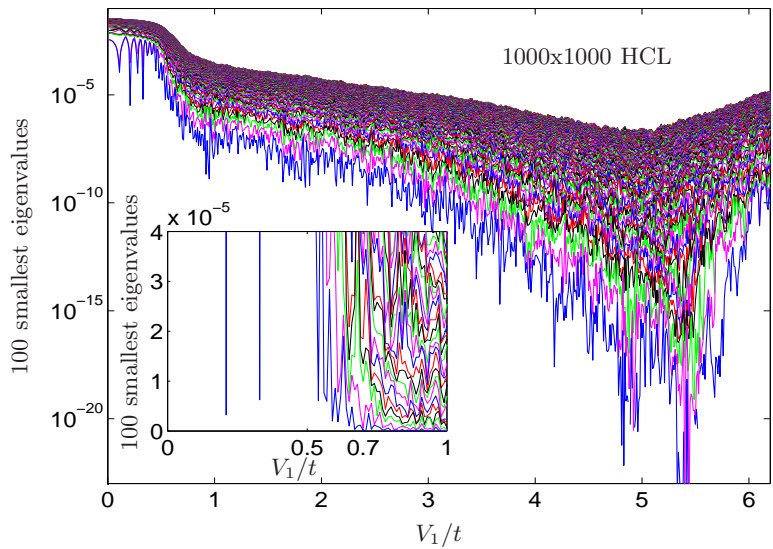}}

\caption{(Color online) The evolution of the lowest 100 eigenvalues above the DP is shown for a $1000\times 1000$ HCL cluster with a given  
Gaussian disorder configuration on a semilogarithmic scale, by changing the strength of the disorder. The inset enlarges the low 
energy structures and the transition from vanishing to diverging behavior. For $V_1>0.7T$, the eigenvalues start to approach zero 
rapidly, as is obvious from the semilogarithmic scale. Their increasing behaviour for $V_1>5$ is due to finite size effects. The 
statistics of the eigenvalues at $V_1=3t$ is depicted in Fig. \ref{hist}} \label{eigen300}
\end{figure}

\begin{figure}[h!]
\centering
\psfrag{x}[t][b][1][0]{$V_1/t$}
\psfrag{y}[b][t][1][0]{100 smallest eigenvalues}
\psfrag{xx}[t][b][0.8][0]{$V_1/t$}
\psfrag{yy}[][][0.8][0]{smallest eigenvalues}

\psfrag{708x708}[][][1][0]{708x708 SQL}
{\includegraphics[width=10cm,height=7cm]{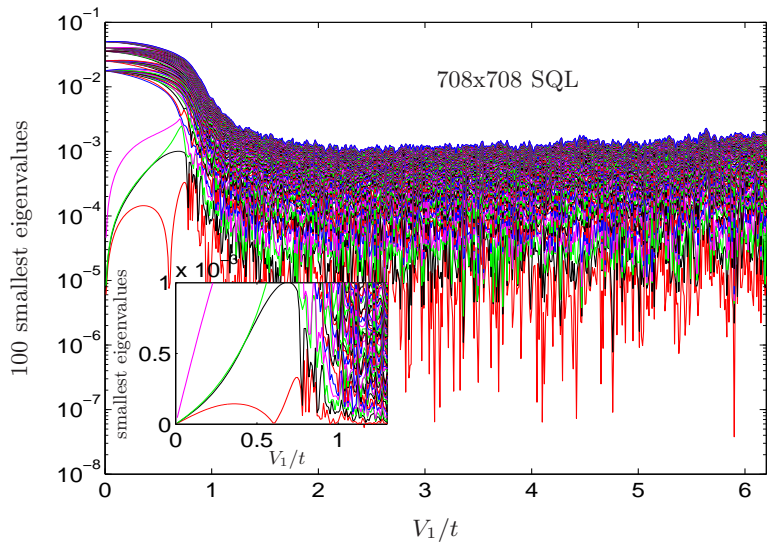}}

\caption{(Color online) The evolution of the lowest 100 eigenvalues above the DP is shown for a $708\times708$ SQL cluster with a given
Gaussian disorder configuration on a semilogarithmic scale, by changing the strength of the disorder. The inset enlarges the low
energy structures and the transition from vanishing to diverging behavior. As opposed the the HCL, the structure of the eigenvalues 
hardly changes for $V_1>t$. Their distribution is shown in Fig. \ref{hist}.}
\label{eigen424sq}
\end{figure}

The investigation of the lowest eigenvalues in the case of unidirectional bond disorder, determining the 
residual DOS, may reveal some structures which are responsible for the aforementioned behavior of the DOS near the DP.
Therefore, we take a single disorder realization of $H_{dis}$, chosen randomly according to a Gaussian distribution.
Then we diagonalize $H_{HCL}+V_1H_{dis}$, using the Lanczos algorithm, and retain the 200 eigenvalues closest to the 
DP (symmetric to the DP). This procedure is repeated for different values of $V_1$.
The result is shown in Figs. \ref{eigen300} and \ref{eigen424sq} as a function of the disorder strength for a 
$1000\times1000$ cluster on the HCL and a $708\times708$ cluster on the SQL, having almost exactly the same number of states. 
This reveals three different regimes: 

 (i) for weak disorder, the distribution of the eigenvalues is rather dilute and is not 
influenced significantly by disorder.  This can explain the zero residual DOS in this case, where a slight rearrangement 
of the eigenvalues change only the slope of the vanishing DOS. 

 (ii) Around $V_1\sim 0.7t$, the pattern changes drastically for the HCL, where the spectrum becomes very dense close to zero 
energy. It keeps on decreasing monotonically down to zero energy. This behavior is responsible for the peak and a possible 
divergence of the DOS.

 (iii) For strong disorder ($V_1/t\sim 5$), the eigenvalues depart from the DP again. This crossover is related to finite size 
effects, since the characteristic disorder value shifts markedly to higher values with increasing system size. 
This is different for the SQL. At low values of $V_1$, the DOS behaves similarly for the HCL as well as for the SQL, where the 
DOS goes down in a power-law fashion, with decreasing exponent, but retains a finite value at the DP. For $V_1>t$, however, 
the eigenvalue pattern is strongly affected only on the HCL by the explicit value of the 
disorder. A direct study of the DOS reveals no peak around the DP for the SQL but a finite 
residual value. This reflects the upper bound which was derived in Sect. 2.3.

\begin{figure}[h!]
\centering
\psfrag{x}[t][b][1][0]{$\omega/t$}
\psfrag{y}[b][t][1][0]{$N(\omega)/\omega\sim\rho (\omega)$}
{\includegraphics[width=7cm,height=7cm]{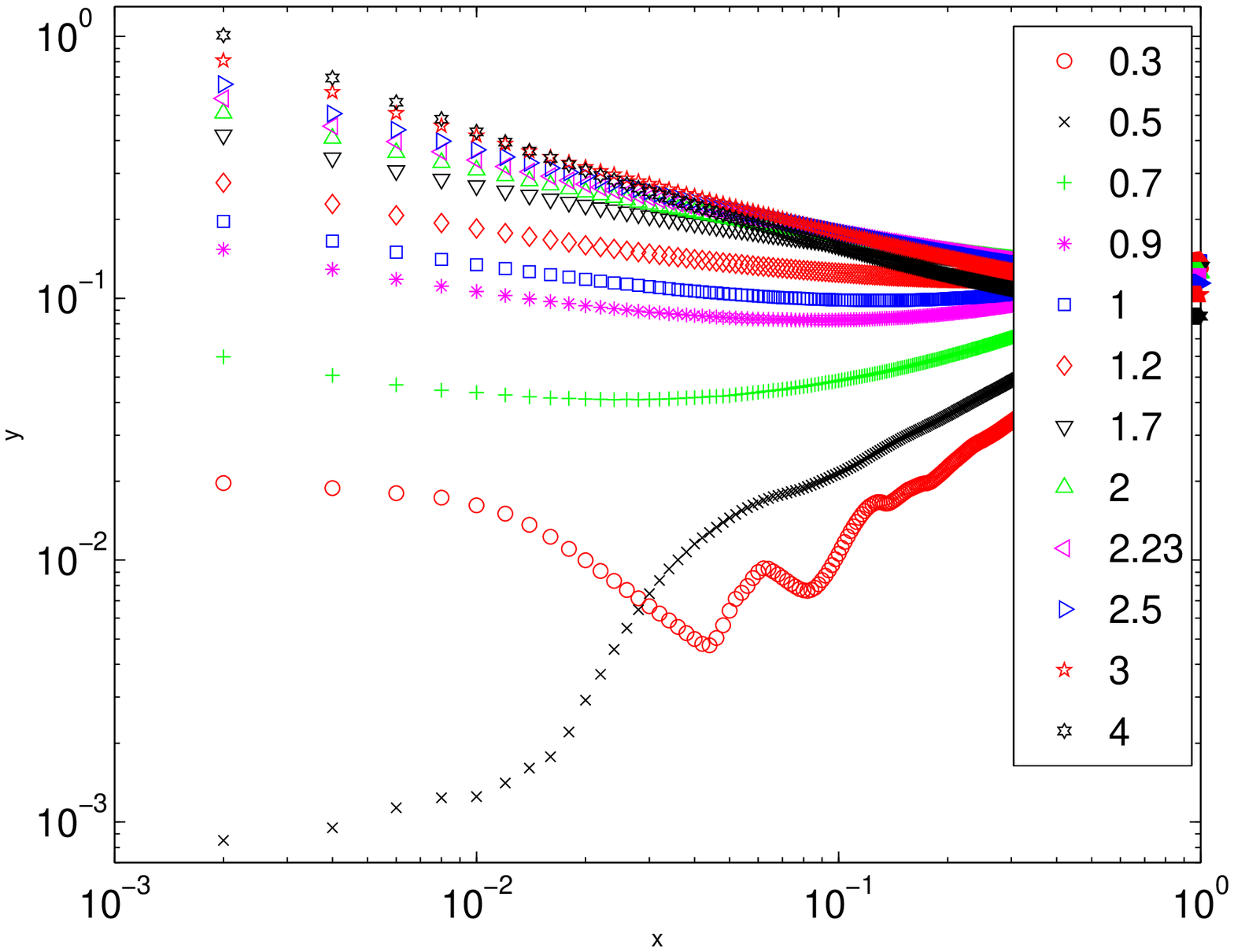}}
\hspace*{5mm}
\psfrag{x}[t][b][1][0]{$\omega/t$}
\psfrag{y}[b][t][1][0]{$N(\omega)/\omega\sim \rho(\omega)$}
{\includegraphics[width=7cm,height=7cm]{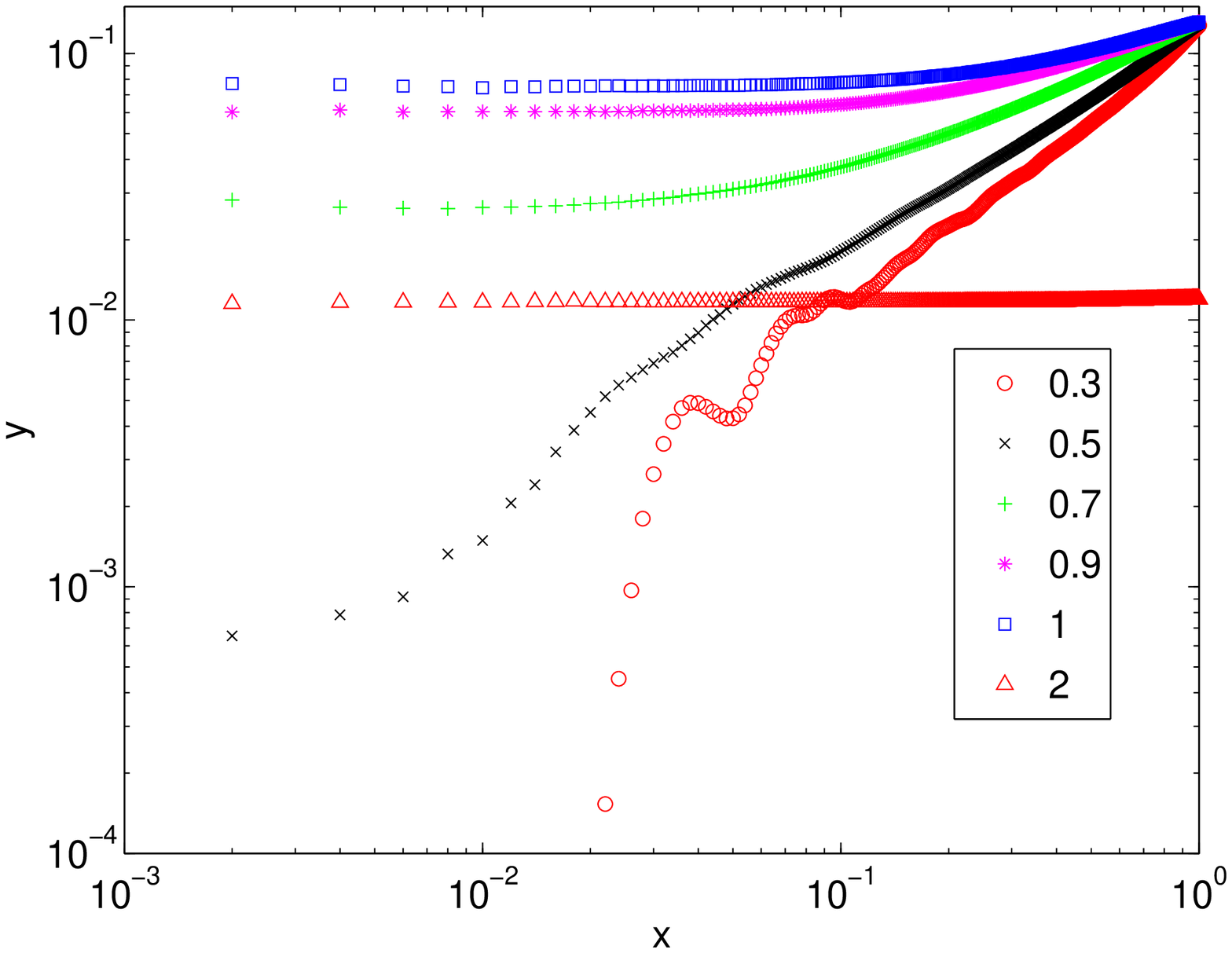}}

\caption{(Color online) The number of states divided by energy ($\sim\rho(\omega)$) is plotted as obtained by exact diagonalization 
on 100x100 honeycomb clusters with Gaussian
unidirectional bond disorder (left panel) and potential disorder (right panel) after 1000 averages for several values of 
the disorder. The upturn with decreasing energy for bond disorder is indicative to the diverging DOS at $E=0$ for 
$V/t\gtrsim 0.6$.}
\label{doshuse}
\end{figure}

\begin{figure}[h!]
\centering
\psfrag{x}[t][b][1][0]{$V_1/t$}
\psfrag{y}[b][t][1][0]{exponent ($\alpha$)}
\psfrag{z}[b][t][1][0]{$z$}
{\includegraphics[width=7cm,height=4cm]{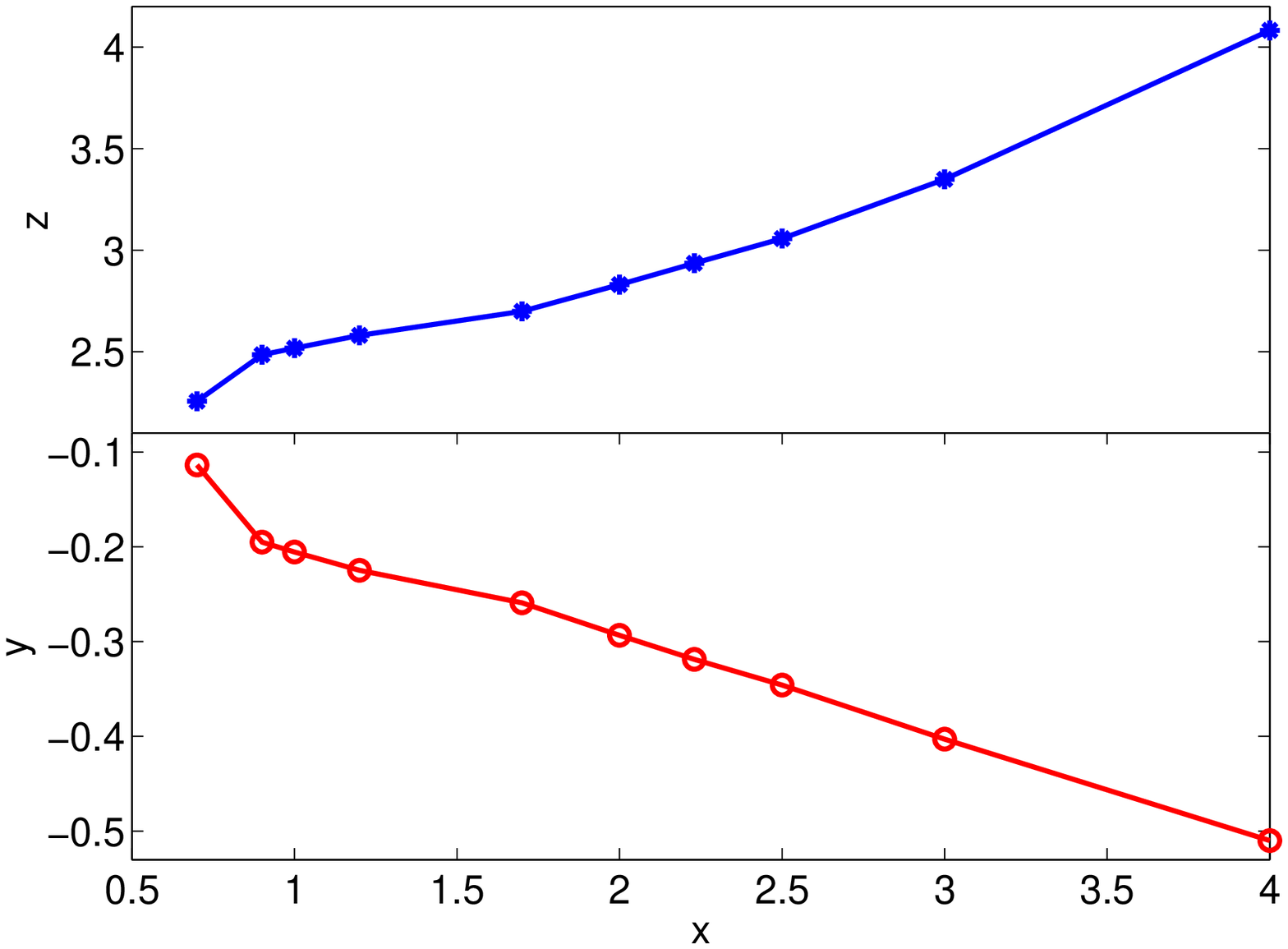}}

\caption{(Color online) The exponents of the DOS ($\rho(\omega)\sim\omega^\alpha$) and the dynamical exponent $z=2/(1+\alpha)$ are
plotted for the HCL for strong disorder. Note the horizontal axis, which is the standard deviation and not the variance.}
\label{expon}
\end{figure}

In order to obtain the DOS, we employ another approach for evaluating this quantity at the DP, which was introduced in Ref. 
\onlinecite{huse}:
We determine the number of states $N(E)$ in a given energy interval $E$ around the Dirac point 
and define the DOS as $\lim_{E\rightarrow 0}N(E)/E$.
As is seen in Fig. \ref{doshuse}, the resulting DOS for $V_1\geq 0.7t$ shows an upturn with decreasing energy for bond disorder, 
which may be indicative 
for a diverging nature of the DOS. The DOS for $V_1=0.5t$ still goes to zero, but  the 0.7 data increases monotonically with 
decreasing energy. This supports the picture, that the residual DOS is indeed zero for $V\lesssim 0.6...0.7t$, and changes to a 
diverging behavior afterwards. The results for $V=0.3t$ are probably strongly affected by finite size effects.
By fitting the resulting curves with a power law, we determine the exponents ($\alpha$) which is  characterizing the DOS close to the DP
(cf. Fig. \ref{expon}). From $\alpha$ the dynamical exponent $z$ follows as $z=2/(1+\alpha)$. 
According to Ref. \cite{huse}, the latter changes its behavior at $z=3$, which is reached here at $V_1/t\sim 2.5$, 
and it increases linearly with $V_1$.
For comparison, the case of potential disorder is plotted as well, where the DOS tends smoothly to a 
constant value at $E=0$. The SQL with $V_1$ disorder exhibits qualitatively similar behavior to the potential disorder case on the 
HCL.

\begin{figure}[h!]
\centering
\psfrag{x}[t][b][1][0]{$E_i$}
\psfrag{y}[b][t][1][0]{distribution of eigenvalues}
\psfrag{xx}[t][b][0.8][0]{$E_i$}
\psfrag{yy}[][][0.8][0]{distribution of eigenvalues}
\psfrag{V}[][][0.8][0]{$V/t=3$}

\psfrag{424x424}[][][1][0]{424x424 SQL}
\psfrag{600x600}[][][1][0]{600x600 HCL}
{\includegraphics[width=7cm,height=7cm]{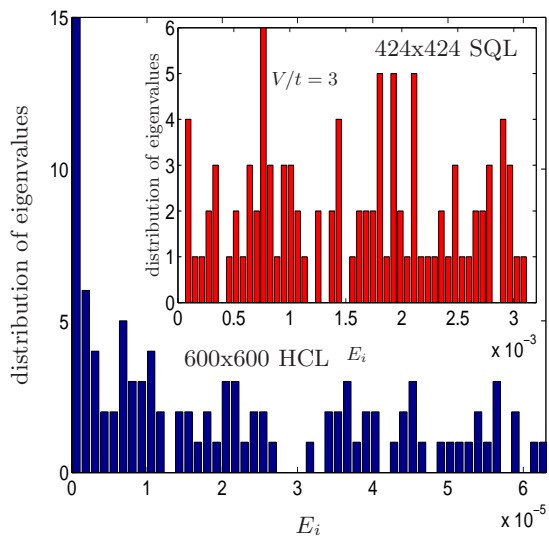}}

\caption{(Color online) A typical distribution of the lowest eigenvalues is shown for $V/t=3$ for both, the HCL and the SQL.
In the former case, the eigenvalues precipitate to zero very fast, resulting in a sharp peak around zero energy. As opposed to this, 
the distribution for the SQL is more uniform, yielding a nondiverging constant DOS.}
\label{hist}
\end{figure}

\begin{figure}[h!]
\centering
\psfrag{x}[t][b][1][0]{$\omega/t$}
\psfrag{y}[b][t][1][0]{$N(\omega)/\omega\sim\rho (\omega)$}
{\includegraphics[width=7cm,height=7cm]{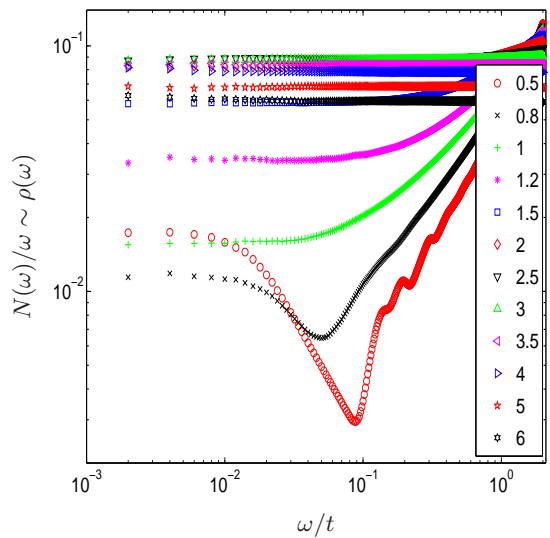}}

\caption{(Color online) The number of states divided by energy ($\sim\rho(\omega)$) is plotted as obtained by exact diagonalization 
on 90x90 square lattice after 1000 averages for several values of the $V_1$ disorder. It resembles closely to the potential disorder 
case of the HCL.}
\label{dossql}
\end{figure}

\subsubsection{Finite DOS on the SQL}

\begin{figure}[h!]
\centering

\psfrag{xx}[t][b][1][0]{$\omega/t$}
\psfrag{yy}[b][t][1][0]{$t\rho(\omega)$}
%{\includegraphics[width=7cm,height=7cm]{dossql90.eps}}
%\hspace*{10mm}
\psfrag{x}[t][b][1][0]{$V_1/t$}
\psfrag{y}[b][t][1][0]{$1/t\rho(0)$}
{\includegraphics[width=7cm,height=7cm]{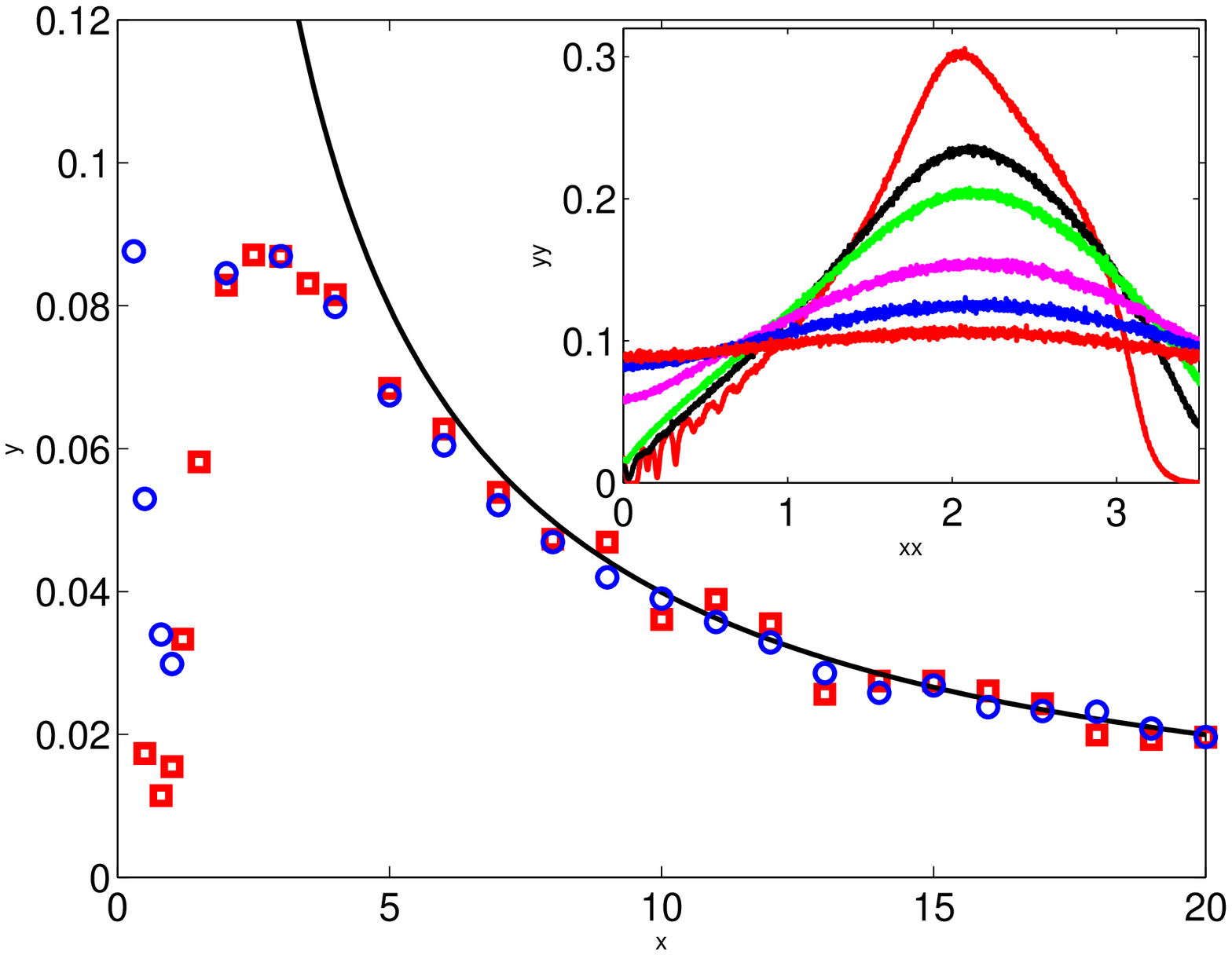}}

\caption{(Color online) The residual DOS of the square lattice, Eq. \ref{ham5} is plotted obtained on 
$90\times90$ and $30\times 30$ clusters with Gaussian
$V_1$ disorder (red squares and blue circles) after 10$^3$ and 10$^4$ averages, respectively. The black 
straight line is 
the upper bound, 
$\rho(0)<1/\sqrt{2\pi}V_1$. Note, that it does not involve any fitting parameter, and becomes very sharp for strong disorder.
Larger systems show similar behavior.
Inset: the DOS of a 90x90 SQL is shown after 1000 averages for $V_1/t$=0.5, 0.8, 1, 1.5, 2 and 2.5 with
decreasing peak-position at $\omega=2t$. 
}
\label{sqresdos}
\end{figure}

Now we turn our attention to the square lattice model in Eq. (5). For the pure system, there is no difference between the HCL and the SQL 
for the DOS near the DP, since excitations close to half filling are Dirac fermions in both cases. Thus, the DOS increases linearly 
with energy. It also exhibits a weak logarithmic singularity at the saddle point of the spectrum, and falls off monotonically with increasing energy 
towards the band edge, as is seen in the inset of Fig. \ref{sqresdos}.
The $V_1$ disorder in Eq. \ref{ham5} on the lattice model plays the role of a random vector potential, which is perpendicular to the (pseudo) spin 
quantization axis $\sigma_3$.
In the presence of $V_1$ disorder %, although it reduces to Dirac fermions with random vector potential in the continuum limit, 
the DOS on the SQL is different from that of the 
HCL with unidirectional bond disorder:  no peak develops at zero energy for strong disorder, and the DOS terminates at a finite value 
with vanishing slope, similarly to potential disorder in the HCL.  
Using an energy window of $t/500$ as for the HCL, we can evaluate the DOS as described above.
The residual values are plotted in Fig. \ref{sqresdos} and compared with the upper bound. As is seen, the upper bound becomes very 
sharp for strong disorder in this case, and does not seem to apply to the HCL with a possibly diverging DOS.

%However, the possible divergent behavior cannot be fitted by a power law of Eq. (\ref{dos01}) with negative exponent.

\section{Conclusions}

We have evaluated the eigenvalues and the average DOS for the tight-binding model on the honeycomb lattice with random
unidirectional bonds and for Dirac fermions on the square lattice with random vector potential. Both models have the same continuum
limit, namely Dirac fermions with a random vector potential. However, in their lattice form they differ substantially
near the Dirac point: In the model on the honeycomb lattice the average DOS has a sharp peak which is not present in the model 
on the square lattice. Although it is not entirely clear, whether or not this peak survives the limit of the infinite system,
its existence on  the finite cluster is remarkable. The evolution of the eigenvalues close to the Dirac point in large systems 
supports the idea of a diverging peak in the DOS.

We have studied the effect of potential disorder on the honeycomb lattice as well, which exhibits qualitatively 
similar behaviour to the square lattice with random vector potential, but differs from the case 
of random
unidirectional bond on the honeycomb lattice at low energies in the DOS: the residual DOS always takes a finite, although 
exponentially small value.
These results can surprisingly well be  reproduced for weak and moderately strong disorder using the self-consistent 
non-crossing approximation, 
expect for the low energy structures in the case of bond disorder.

Using the mapping of the model on the HCL to the $SU(2)$ gauge field
theory \cite{nersesyan94,caux98,bhaseen01,foster08}, the presumably exact power law of the latter $\rho\sim |E|^{1/7}$ 
of Ref. \cite{nersesyan94} represents a puzzle for the approximation of disordered lattice models by their corresponding continuum
counterparts. The same is true for the model on the square lattice, where the DOS at the Dirac point has an upper bound according to
Eq. (\ref{upperbound}). In contrast, for the continuum limit several groups found a power law with the exponent 
\cite{ludwig94,nersesyan94}
\[
\alpha =\frac{1-g/\pi}{1+g/\pi}
\]   
which is negative for sufficiently strong disorder.
This poses severe questions on the applicability of universality idea.
Although both lattice models reduce to the same continuum limit and are expected to behave in a similar manner, as dictated by 
the common continuum limit, this is apparently not the case here.
We have also checked the case of uniform disorder distribution, and found similar results.
The above results were found to be robust with respect to variations of system size, boundary conditions, and disorder distribution.

%\begin{acknowledgments}
We acknowledge useful discussions with O. Vafek, I. Herbut and R. Moessner. We are grateful to B. Schmidt and A. 
V\'anyolos for technical assistance.
This work was supported by the Hungarian
Scientific Research Fund under grant number OTKA K72613, by a grant from the Deutsche Forschungsgemeinschaft  
and in part by the Swedish Research Council.
%\end{acknowledgments}

\appendix

\section{Discrete symmetry}

From $h^T_j=-h_j$ and $\sigma_1^T=\sigma_1$, $\sigma_2^T=-\sigma_2$ follows 
\[
H^T=(-h_1+V_1)\sigma_1-(-h_2+V_2)\sigma_2 \ .
\]
Next, $D$ changes the sign of nearest-neighbor matrix elements:
\[
DH^TD=(h_1+V_1)\sigma_1-(h_2+V_2)\sigma_2 \ ,
\]
and $\sigma_1$ anticommutes with $\sigma_2$:
\[
\sigma_1 DH^TD\sigma_1 =(h_1+V_1)\sigma_1+(h_2+V_2)\sigma_2=H \ .
\]

\section{Matrix elements of the Green's function}

The spatial diagonal matrix elements of the Green's function
have been given in Eq. (\ref{expansa}).
Another way to write $G_{rr}$ is by projecting it with $P_r$ onto the site $r$.
This gives the matrix identity \cite{blockmatrix}
\[
G_{rr}\equiv P_rGP_r
=[i\epsilon+V_r\sigma_1+V_r'\sigma_2-P_rH({\bf 1}-P_r)
G_{{\bf 1}-P_r}({\bf 1}-P_r)HP_r]_{P_r}^{-1} \ ,
\]
where $G_{{\bf 1}-P_r}$ is the Green's function $G(i\epsilon)=(i\epsilon+H)^{-1}$ on the
Hilbert space where the site $r$ has been removed. The $2\times2$ matrix $P_rH({\bf 1}-P_r)
G_{{\bf 1}-P_r}({\bf 1}-P_r)HP_r$ does not depend on the random variables
$V_r$ and $V_r'$. Its general form is
\[
P_rH({\bf 1}-P_r)G_{{\bf 1}-P_r}({\bf 1}-P_r)HP_r=-\left[\begin{matrix}
iX_0 +X_3 & -iX_2+X_1 \cr
iX_2+X_1 & iX_0 -X_3\cr
\end{matrix}\right] \ .
\]
Therefore, $G_{rr}$ reads
\[
G_{rr}=\left[\begin{matrix}
i\epsilon+iX_0 +X_3 & -iX_2+X_1+V_r-iV_r' \cr
iX_2+X_1+V_r+iV_r' & i\epsilon+iX_0 -X_3\cr
\end{matrix}\right]^{-1}
\]
\beq
=-{1\over (\epsilon+X_0)^2+X_3^2+(X_1+V_r)^2+(X_2+V_r')^2}
\left[\begin{matrix}
i\epsilon+iX_0 -X_3 & iX_2+X_1+V_r +iV_r'\cr
-iX_2+X_1+V_r-iV_r' & i\epsilon+iX_0 +X_3 \cr
\end{matrix}\right] \ .
\label{gf1}
\eeq
This result can be compared with Eq. (\ref{expansa}) to obtain the relations
\beq
X_1=-V_r+\frac{g_1}{-g_0^2+g_1^2+g_2^2} \ ,
\hskip0.3cm
iX_0=-i\epsilon-\frac{g_0}{-g_0^2+g_1^2+g_2^2} \ ,
\hskip0.3cm
X_2=-V_r'+\frac{g_2}{-g_0^2+g_1^2+g_2^2}
%g_2={X_2\over (\epsilon+X_0)^2+X^2+X_2^2} 
\ ,
\label{xparameters}
\eeq
and
\[
X_3=0 \ .
\]
All three matrix elements $X_0,X_1,X_2$ are real, since $g_0$ is purely imaginary, and
$g_1$ as well as $g_2$ are real.

Finally, we can use the block-matrix inverse to show that $g_0$ is proportional to
$-i\epsilon$ with a positive proportionality factor. Choosing the diagonal blocks
with respect to the sublattice (or spinor) index $j$, we obtain
\[
G_{11}=[i\epsilon-(h_1+V_1-ih_2-iV_2)(h_1+V_1+ih_2+iV_2)/i\epsilon)]^{-1}
=-i\epsilon [\epsilon^2+(h_1+V_1-ih_2-iV_2)(h_1+V_1+ih_2+iV_2)]^{-1}
\]
\[
=-i\epsilon [\epsilon^2+(h_1+V_1-ih_2-iV_2)(h_1+V_1-ih_2-iV_2)^\dagger]^{-1}
\]
and
\[
G_{22}=[i\epsilon-(h_1+V_1+ih_2+iV_2)(h_1+V_1-ih_2-iV_2)/i\epsilon)]^{-1}
=-i\epsilon [\epsilon^2+(h_1+V_1+ih_2+iV_2)(h_1+V_1-ih_2-iV_2)]^{-1}
\]
\[
=-i\epsilon [\epsilon^2+(h_1+V_1-ih_2-iV_2)^\dagger(h_1+V_1-ih_2-iV_2)]^{-1} \ .
\]
Thus $iX_0+i\epsilon=ci\epsilon$ with $c>0$.

\end{document}